# Search for ultra high energy cosmic ray sources. FRI radio galaxy Centaurus A

**Oleksandr Sushchov[1], Oleh Kobzar[1], Bohdan Hnatyk[2], Volodymyr Marchenko[1]**

[1] Astronomical Scientific Research Center of Chernihiv National Pedagogical University
53, Hetman Polubotok Str., Chernihiv, 14013, Ukraine

[2] Astronomical Observatory of Taras Shevchenko Kyiv National University
3, Observatorna Str., Kyiv, 04053, Ukraine

volodymyr.marchenko@gmail.com

**Abstract.** The propagation of ultra high energy cosmic rays in Galactic and extragalactic magnetic fields is investigated in the present paper. The motion of charged particles of different energies and chemical composition is simulated using different Galactic magnetic field models. Positions for the real sources of events registered at the Auger observatory are calculated taking into account the influence of both Galactic and extragalactic turbulent fields. The possibility of their correlation with the Centaurus A radio galaxy is analyzed.

## INTRODUCTION

Cosmic rays (CR) are known as fluxes of high-energy subatomic particles, photons or neutrino generating extended atmospheric showers of secondary particles that interact with molecules of nitrogen and oxygen which are prevalent in the Earth atmosphere's upper strata. CR possessing the energy $E > 10^{19}$ eV arrive at the Earth with the interval of less than one event per year over 1 km$^2$ in $\pi$ steradian (i.e. with the energy flux of 30 eV/cm$^2\cdot$sec) [30].

There is no specific criterion of identifying CR energy as ultra high. As a rule it is energy over $10^{19}$ eV. Ultra high energy cosmic rays (UHECR) are believed to be of extra-galactic origin due to the absence of sources powerful enough to provide their sufficient acceleration within our Galaxy as well as due to almost isotropic large-scale distribution of CR along the lines of their entering the atmosphere [3].

The hypothesis concerning UHECR's astrophysical nature is also supported by registering the Greisen-Zatsepin-Kuzmin effect [15, 34] (the so called GZK-cutoff of CR's energy spectrum) in the HiRes experiment [1], as well as in observations carried out at the Auger observatory [2].

On the way from their sources towards the Solar system CR lose their energy due to the redshift as well as forming pions and electron-positron pairs while interacting with relict photons. These processes define the CR horizon – the typical distance from the Earth that can be covered by CR possessing the final energy that surpasses a certain value. Thus 70% of possible sources of protons and iron nuclei with the energy of $E > 10^{20}$ eV should be located within 70 Mpc from the Earth while the source of CR with the energy of $E > 6\cdot10^{19}$ eV are supposed to be found within 250 Mpc [19].

The correlation between CR and galaxies from the active galactic nuclei (AGN) Veron-Cetti-Veron (VCV) catalogue [33] can be accepted as a possible explanation of UHECR's nature, provided their origin is extragalactic while their horizon is energy dependent which agrees with the data on GZK-cutoff. Yet the empiric data regarding the said correlation is insufficient for final acknowledging AGN as UHECR's sources. The AGN's distribution is commensurable to that of regular galaxies (i.e. the distribution of matter in general) therefore they may appear to be only the indicators of real sources [10]. Except for AGN, such classes of astrophysical objects as gamma-ray bursts, young magnetars and shock waves in areas of large-scale structure's formation have at times been suggested as accelerators of extragalactic CR. Theoretic calculations require that CR's potential sources should be powerful enough to accelerate the particles up to ultra high energy values. Out of stable astrophysical objects within the red shift $z = 1$ only a few (like Fanaroff-Riley radio galaxies of type I and II) are marked by physical conditions favorable for accelerating protons up to ultra high energy. However, these conditions may turn real for certain classes of variable sources (those with a period of activity shorter than the temporal delay of CR movement the latter being caused by magnetic fields on the way of their propagation) like AGN flares, young magnetars or gamma-ray bursts. As heavy nuclei are concerned, the scope of their potential sources is wider for they can be accelerated by less powerful astrophysical objects. If this case proves to be true, several closest radio galaxies or even a sole galaxy of Centaurus A type could be regarded as contributors to the observed UHECR flux. Still, “radio quiet” AGN and gamma-ray bursts are on the list too [22].

The energy spectrum observed on the surface of the Earth is not identical to the initial spectrum of CR generation. It has been impacted by Galactic and extragalactic magnetic fields and CR's interaction with the microwave background. Unlike protons, heavy nuclei can be accelerated to higher energy values which may result into gradual transformation in CR's composition – from proton to a heavier (up to iron) type – in the high-energy part of spectrum [13]. The so called "ankle" (the decrease in spectrum's steepness around $5 \cdot 10^{18}$ eV) can be interpreted as either an illustration of UHECR's sources' transfer from Galactic to extragalactic once at certain energy level or as a distortion of CR's mostly proton spectrum due to emergence of $e^+ e^-$ pairs when protons interact with relict photons [5].

The analysis of CR's chemical composition carried out by the Auger observatory testifies to the fact that beginning from $10^{19}$ eV energies there occurs a shift towards heavy nuclei (starting at $3 \cdot 10^{19}$ eV). The data by Yakutsk EAS Array [14] on muon component of atmospheric showers also prove the shift in UHECR's chemical composition towards heavy nuclei while the HiRes measurements and preliminary data by Telescope Array [18] come in accord with CR's predominantly proton composition. The reason of the discrepancy in the primary experiments' results is still to be discussed.

Anisotropy of UHECR's arrival directions and their correlation with possible sources, AGN in particular, energy spectrum and chemical composition are rather promising research areas that aim at explaining UHECR's nature and origin. If the registered decrease in CR flow is in fact triggered by the GZK effect, then there is a corresponding GZK horizon at approximately 100 Mpc; thus CR with the initial energy of about $10^{20}$ eV coming from farther distances will be registered as those possessing considerably lesser energy. As the distribution of matter in the circumference of 100 Mpc within our Galaxy is not isotropic, we have a chance of finding anisotropy in the registered data [31].

According to the analysis of the refreshed data from the Auger observatory [4], the registered UHECR's correlation with the galaxies from the VCV catalogue has diminished from level $3\sigma$ to $2\sigma$ compared to the previous version of data update. As the result only 30% of UHECR potentially correlate with the directions towards AGN whereas the rest manifest the signs of isotropic distribution. The only exception is the neighborhood of the closest to the Earth active galaxy Centaurus A where the registered set of ultra energy events appeared to be a lot more volumetric than it could be statistically correct to expect.

In this paper we verify the possibility of the observed in the area of Centaurus A events being UHECR accelerated in this very galaxy. Therefore we solve the reversed task by modeling CR's flow trajectory on the basis of data concerning the said galaxy, acquired at the Auger facility. The model takes into account the influence of Galactic and extragalactic magnetic fields as well as the CR chemical composition.

# MODELING

Magnetic field distorts the CR's charged particles' trajectory via Lorenz force. If the field is static it does not affect the particle's energy. Considering the fact that typical values of CR energy far exceed the particles' rest energy we assume that they propagate with velocity close to the speed of light. In this case the equations describing the motion of ultra-relativistic particles in the magnetic field $\boldsymbol{B}(\mathbf{r})$ are:

$$\frac{d\mathbf{v}}{dt} = \frac{qc^2}{E}[\mathbf{v} \times \mathbf{B}], \qquad \frac{d\mathbf{r}}{dt} = \mathbf{v},$$

where $q$ is particle's charge, $E$ — its energy, provided the Lorenz factor $\gamma >> 1$ and velocity absolute value $|\mathbf{v}| = c\sqrt{1 - \frac{1}{\gamma^2}} \approx c$.

Considering the complex structure of the magnetic field solving these equations analytically is impossible. Therefore trajectory calculation was carried out via numeric modeling.

# MAGNETIC FIELDS

Modeling the motion of UHECR we considered influence of Galactic as well as extragalactic magnetic fields. Galactic magnetic field consists of regular and random components. The regular component's structure is believed to generally follow the matter's distribution in the Galaxy [16, 17]. Nowadays the source and structure of extragalactic magnetic field are not exactly clear. Thus while solving specific tasks it is defined as having random structure [8].

## *Galactic magnetic field*

***A.*** *Regular component*. There are several models describing regular Galactic magnetic field [29]. They differ in both parameters' numeric values and presence and structure of the field's components. The regular component of Galactic magnetic field is rather conveniently described via the spiral structure of 2π-symmetry (axisymmetric spiral (ASS)) or π- symmetry (bisymmetric spiral (BSS)) [28]. In our research we have applied the most recent models [23, 20, 24]. They present the magnetic field as a superposition of the disc component and the field of Galactic halo. In [23] and [20] BSS symmetry is used for describing the disc field whereas in [24] both ASS and BSS disc field's symmetry types are considered (henceforth we treat them off as different models).

The disc field comprises radial and azimuth components which are set in cylindrical coordinates in the disc's area by expressions

$$B_r = B(r,\theta)\sin p\,,$$

$$B_\theta = B(r,\theta)\cos p\,,$$

where pitch angle $p$ is the angle between the magnetic vector at a certain point and the normal to radius-vector **r** in this point.

The function $B(r,\theta)$ is set by the equation of logarithm spiral:

$$B(r,\theta) = B(r)\cos\left[\theta - \frac{1}{\tan p}\ln\left(\frac{r}{\xi_0}\right)\right], \qquad (1)$$

or

$$B(r,\theta) = B(r)\cos\left[\theta - \frac{1}{\tan p}\ln\left(\frac{r}{R_\odot}\right) + \varphi\right], \qquad (2)$$

Parameters in formulae (1) and (2) are set by expressions

$$\varphi = \frac{1}{\tan p}\ln\left(1 + \frac{d}{R_\odot}\right) - \frac{\pi}{2},$$

$$\xi_0 = \left(R_\odot + d\right)\exp\left(-\frac{\pi}{2}\tan p\right),$$

where $R_\odot = 8.5\,\mathrm{kpc}$ is the distance from the Galactic center to the Solar system, $d$ is the distance from the Solar system to the closest field's inversion point.

The function of the radial profile $B(r)$ is set by

$$B(r) = \begin{cases} B_\odot \dfrac{R_\odot}{r\cos\varphi} = B_0 \dfrac{R_\odot}{r}, & r > R_\mathrm{C}, \\ B_\odot \dfrac{R_\odot}{R_\mathrm{C}\cos\varphi} = B_0 \dfrac{R_\odot}{R_\mathrm{C}}, & r < R_\mathrm{C}, \end{cases}$$

where $B_\odot$ is local field near the Solar system.

The vertical profile of the disc field above the Galactic plane and under it is considered exponentially decreasing:

$$B(r,\theta,z) = B(r,\theta)\exp\left(-\frac{|z|}{z_0}\right).$$

The differences between the used models mostly consider the choice of the field's shape — (1) or (2) and the values of some parameters as reflected in Table 1.

**Table1.** Parameters of Galactic magnetic field's disc component

| Field model | | $p$, degrees | $R_C$, kpc | $d$, kpc | $\xi_0$, kpc | $z_0$, kpc | $B_\odot$, μG | $B_0$, μG |
|---|---|---|---|---|---|---|---|---|
| [23] | | –10 | 4.0 | — | 9.0 | 1.0 | — | 3.0 |
| [20] | | –8 | 4.0 | –0.5 | — | 0.2 | 2.0 | — |
| [24] | ASS | –5 | 5.0 | –0.6 | — | 1.0 | 2.0 | — |
| | BSS | –6 | 5.0 | –0.6 | — | 1.0 | 2.0 | — |

In models [23] and [20] the field of Galactic halo comprises poloidal and toroidal components while model [24] contains the toroidal component only. For the description of the toroidal field we use the model of discs located above and under the Galactic plane. The toroidal field's parameters are set by expressions

$$B_x = -B_T \mathrm{sign}(z)\left[1+\left(\frac{|z|-h}{w}\right)^2\right]^{-1}\cos\theta,$$

$$B_y = B_T \mathrm{sign}(z)\left[1+\left(\frac{|z|-h}{w}\right)^2\right]^{-1}\sin\theta,$$

where $h$ is the height of discs above and under the Galactic plane, $\omega$ is half-width of Lorenz distribution.

The function $B_T$ in model [23] is given by the following expression

$$B_T = B_{T\max}\left[\Theta(R_T - r) + \Theta(r - R_T)\exp\left(-\frac{r}{R_T}\right)\right],$$

while in model [20]

$$B_T = B_{T\max}\left[\Theta(R_T - r) + \Theta(r - R_T)\exp\left(\frac{R_T - r}{R_T}\right)\right],$$

where $\Theta$ is Heaviside function, $R_T$ is toroid's characteristic radius.

In model [24]

$$B_T = B_{T\max}\frac{r}{R_T}\exp\left(\frac{R_T - r}{R_T}\right).$$

The field's dipole component is described by standard equations:

$$B_x = -3\mu_G \cos\phi \sin\phi \sin\theta / \rho^3 ,$$

$$B_y = -3\mu_G \cos\phi \sin\phi \cos\theta / \rho^3 ,$$

$$B_z = \mu_G \left(1 - 3\cos^2\phi\right) / \rho^3 ,$$

where $\rho \equiv \sqrt{r^2 + z^2}$ , $\cos\phi \equiv z/\rho$, $\sin\phi \equiv r/\rho$, $\mu_G$ is the magnetic dipole momentum.

The parameters' numeric values in the chosen models are provided in Table 2.

***B.*** *Random component*. It is assumed [6] that Galactic magnetic field's random component's impact primarily results into widening the range of UHECR's possible arrival directions relative to the direction defined by the deflection of the trajectory in the regular field. In this case the real location of CR's source is not explicated. Furthermore, under certain conditions the so called “lensing effect” in the magnetic field may occur and generate several images of CR's source [13]. Yet studying this kind of impact may be fruitful for exploring the properties of Galactic magnetic field and CR's propagation.

CR's ultra high energy is marked by the value of Larmor radius that by far exceeds the length of field's coherence $l_0$, the latter understood as the distance at which the field's random re-orientation occurs. Thus to estimate the effect caused by the random magnetic field it is sufficient to consider two parameters: $l_0$ and field's magnitude $B_{rms}$ [12]. The field $B_{rms}$ is characterized by exponentially decreasing vertical profile $B_{rms} = B_0 \exp\left(-\frac{|z|}{z_0}\right)$ [11].

**Table 2.** Parameters of Galactic halo's magnetic field

<table>
<tr><th colspan="2">Field model</th><th>$R_T$,<br>kpc</th><th>$h$,<br>kpc</th><th colspan="2">$w$,<br>kpc</th><th colspan="2">$B_{Tmax}$,<br>μG</th><th>$\mu_G$,<br>$\mu G \cdot kpc^3$</th></tr>
<tr><td colspan="2">[23]</td><td>15.0</td><td>1.5</td><td colspan="2">0.3</td><td colspan="2">1.0</td><td>100</td></tr>
<tr><td colspan="2">[20]</td><td>8.5</td><td>1.5</td><td colspan="2">0.3</td><td colspan="2">1.0</td><td>123</td></tr>
<tr><td rowspan="2">[24]</td><td>ASS</td><td>8.0</td><td>1.3</td><td>0.25<br>(inner)</td><td>0.4<br>(outer)</td><td>4.0<br>(North)</td><td>2.0<br>(South)</td><td>—</td></tr>
<tr><td>BSS</td><td>8.0</td><td>1.3</td><td>0.25<br>(inner)</td><td>0.4<br>(outer)</td><td>4.0<br>(North)</td><td>4.0<br>(South)</td><td>—</td></tr>
</table>

According to the observation data Galactic magnetic field's random component is commensurable to the regular one [23]. In this paper we employ the values $l_0 = 50$ pc, $B_0 = 4$ μG, $z_0 = 3$ kpc [13].

CR deflection $\vartheta$ in the random magnetic field on the covered distance $L$ is set by the following expression [9]

$$\left\langle \vartheta^2 \right\rangle = \frac{2}{9} \cdot \left( \frac{Ze}{E} \cdot c \right)^2 \left\langle B^2 \right\rangle L l_0 , \qquad (3)$$

where $Ze$ is particle's charge, $E$ is its energy.

The distance covered by UHECR (those registered by the Earth-located detectors) in the Galactic turbulent field can be estimated as

$$L_{\mathrm{gal}} = \min\left( \frac{z_0}{\sin b_{\mathrm{G}}};\ L_{\max} = 20\ \mathrm{kpc} \right),$$

where $b_{\mathrm{G}}$ is Galactic latitude of CR's arrival direction. Thus we acquire the value of CR's final deflection

$$\vartheta = 22^\circ \cdot Z \cdot \left( \frac{L_{\mathrm{gal}}}{1\ \mathrm{kpc}} \right)^{1/2} \cdot \left( \frac{E}{10^{18}\ \mathrm{eV}} \right)^{-1} .$$

### *Extragalactic magnetic field*

There are structures in the Universe comprising clusters of galaxies, filaments, layers of increased density and voids with low density. It is assumed that in entities of this kind magnetic field is boosted due to the formation of large scale structures. Diverse numeric modeling of the said process demonstrates correspondence of extragalactic magnetic field's distribution with that of matter [26, 27]. Astrophysical objects, UHECR's sources in particular, are normally located within the structured areas. Thus these magnetic structures as well as the Galactic magnetic field necessarily impact the propagation of CR. The structured extragalactic magnetic field influences both CR's deflection and the time of their reaching the observer.

According to the recent research of the gamma-ray range, extragalactic magnetic field possesses the value of approximately $10^{-15}$ G in the voids [7]. Although this estimation is rather contradictory; the prior estimation of magnetic field's lower limit being $10^{-17}$ — $10^{-15}$ G [31]. In the suggested calculations we employ the simplest model in which space is divided into cubic cells of $l_c$, size. The field is considered uniform within one cell while its direction varies randomly in between the cells. To limit the size of field $B$ we used the value resulting from observation data concerning distant objects' polarization plane's Faraday's rotation [21]:

$$\langle B\rangle\sqrt{l_{0\text{eg}}} \le 10^{-9}\ \text{G}\cdot\text{Mpc}^{1/2},$$

where $l_{0\text{eg}}$ is magnetic field's coherence length. Generally $l_{0\text{eg}}$ does not equal $l_c$ strictly, though this difference is not significant for estimating UHECR's propagation in extragalactic magnetic field.

For random extragalactic magnetic field like in the case of Galactic magnetic field's random component, CR's deflection is calculated through formula (3).

Considering the limitations over the value of extragalactic magnetic field we acquire numeric values of deflection for CR with energy $E$ and charge $Z$, located at the distance $L_0$ from random sources:

$$\vartheta = 25^\circ \cdot Z \cdot \left(\frac{L_0}{1\,\text{Mpc}}\right)^{1/2} \cdot \left(\frac{E}{10^{18}\ \text{eV}}\right)^{-1}.$$

## RESULTS

***Magnetic field's general impact on CR's deflection.*** In the first approximation we modeled CR's movement in the magnetic field considering the regular component only. The employed “back tracking” method comprises the following procedures. Prior to all the charge of the particle corresponding to the registered event was identified. Secondly, we modeled the movement of an anti-particle starting from the Earth in the direction opposite to the registered CR and possessing the same energy. Under those conditions the anti-particle's trajectory reconstructed the trajectory of the real CR. This method allows calculating possible locations of the source of any registered event depending on its chemical composition.

Another approximation was considering the random magnetic field. Its impact results into widening the area of the source's possible location provided its location calculated in the first approximation is unchanged. We can estimate the value of the trajectory's average deflection while the direction of deflection that depends on a specific realization of the random magnetic field remains unidentified. Therefore, lacking specific knowledge of the random component's structure we may apply the “back tracking” method to relatively close objects and CR with low charge for in this case the distortion of the reconstructed trajectories by the random component is not significant.

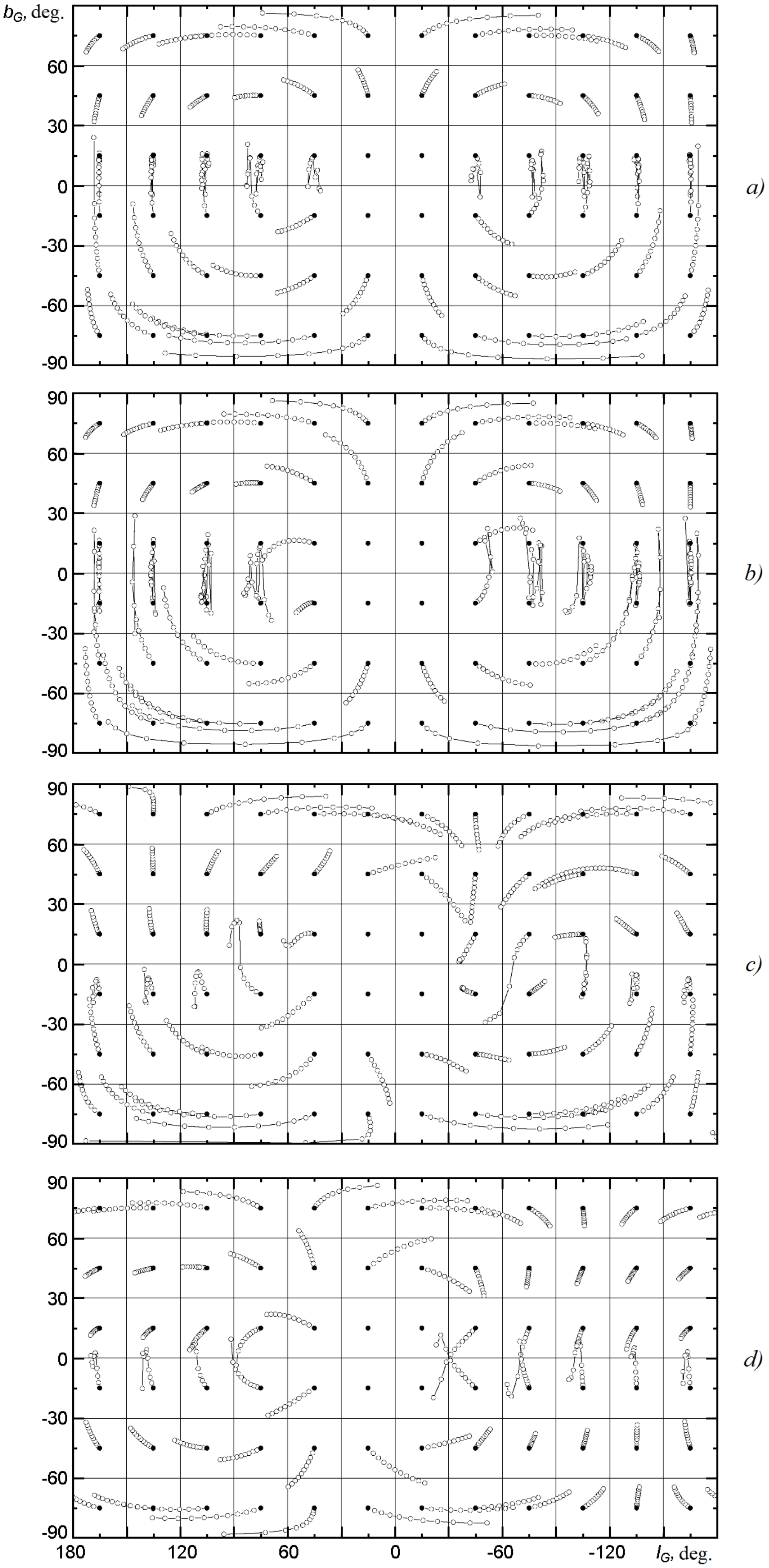

**Figure 1.** CR's deflection in the regular Galactic field calculated according to the models: *a* — model [24], ASS symmetry; *b* — model [24], BSS symmetry; *b* — model [23]; *d* — model [21]

Figure 1 demonstrates the UHECR's trajectory deflection in Galactic coordinates calculated on the basis of various models of the regular Galactic magnetic field. The dots mark the directions where nuclei with the energy of $10^{20}$ eV were launched from the Earth. The circles mark their movement direction in extragalactic space after the deflection in the regular Galactic field. Light dots move away from the dark ones as the launched nuclei's charge grows from 1 to 10.

We have to stress that the image on Figure 1 is distorted especially in the near-pole areas. Therefore in order to define the real angular distance $\varphi$ on the celestial sphere between two points with coordinates ($l_{G1}$; $b_{G1}$) and ($l_{G2}$; $b_{G2}$) we should use the formula

$$\varphi = \arccos\left[\sin b_{G1} \sin b_{G2} + \cos b_{G1} \cos b_{G2} \cos\left(l_{G2} - l_{G1}\right)\right].$$

Comparing the results achieved by using different models we may highlight the following regularities.

1. All models reflect the asymmetry of CR's deflection relative to the Galactic plane. This fact could be explained by different types of symmetry of the field's components relative to this very plane (the disc component is symmetric while the toroidal one is asymmetric).

2. In the area of low latitudes ($|b_G| < 30°$) the field's activity largely depends on a specific model. Obviously in the close vicinity to the Galactic plane the field's disc component causes dominant influence upon the trajectories. Thus they become more sensitive to the models' subtle peculiarities connected with the component's spiral structure, primarily with the choice of the pitch-angle, distance between the Solar system and the closest point of the field's inversion etc.

3. The field's structure around the Galactic center is not precisely known and its value there may by far exceed that at the periphery. Therefore approaching this area dramatically increases the role of the means employed in different models to set up the magnetic field in the central part of the Galaxy so that the observed filaments should be described correctly [23]. Due to vivid non-monotony of the calculated trajectories we have so far left the central area out of consideration for it should be the object of a detailed research.

4. In areas $|b_G| > 30°$ the general pattern of CR's calculated deflections fully agrees with the field's toroidal component. The absence of the dipole component in both models [24] explains the qualitative symmetry relative to the meridian that passes through the way towards the Galactic center (Figure. 1, a, b). The dipole component in models [23] and [20] results into the symmetry area's turning relative to the zero meridian at a certain angle (Figure 1, c, d). In this case the turning angle is somewhat different in the northern and southern parts (Figure 1, c).

5. A number of common regularities noted in the different models of the magnetic field can be explained by the field's similar structure (spiral disc field and Galactic halo's field). The differences are explained by the presence or absence of the dipole component as well as by various parameter values used in the models.

***Centaurus A.*** The Auger observatory registered a set of UHECR in the region of Centaurus A galaxy which is the closest to the Solar system active one. The origin of the registered CR is most likely affiliated with the said galaxy [4]. We have modeled the CR's motion in the magnetic field using the above-described methodology.

Figure 2 demonstrates the results of calculations carried out on the basis of various models of the regular Galactic magnetic field. Circles with figures denote the set of events registered by the Auger facility. Circles with the chemical elements symbols correspond to the calculated locations of UHECR's sources for the indicated particle types. Radii of all circles reflect the Auger detectors' experiment error within the confidence interval of $1\sigma$. Results depicted in Figure 2a, were obtained via the use of model [23]; those in Figure 2b were achieved as the result of using model [20]. The figures also demonstrate the outline of Centaurus A’s radiation areas. These areas are known to have conditions for accelerating CR up to ultra high energies [25].

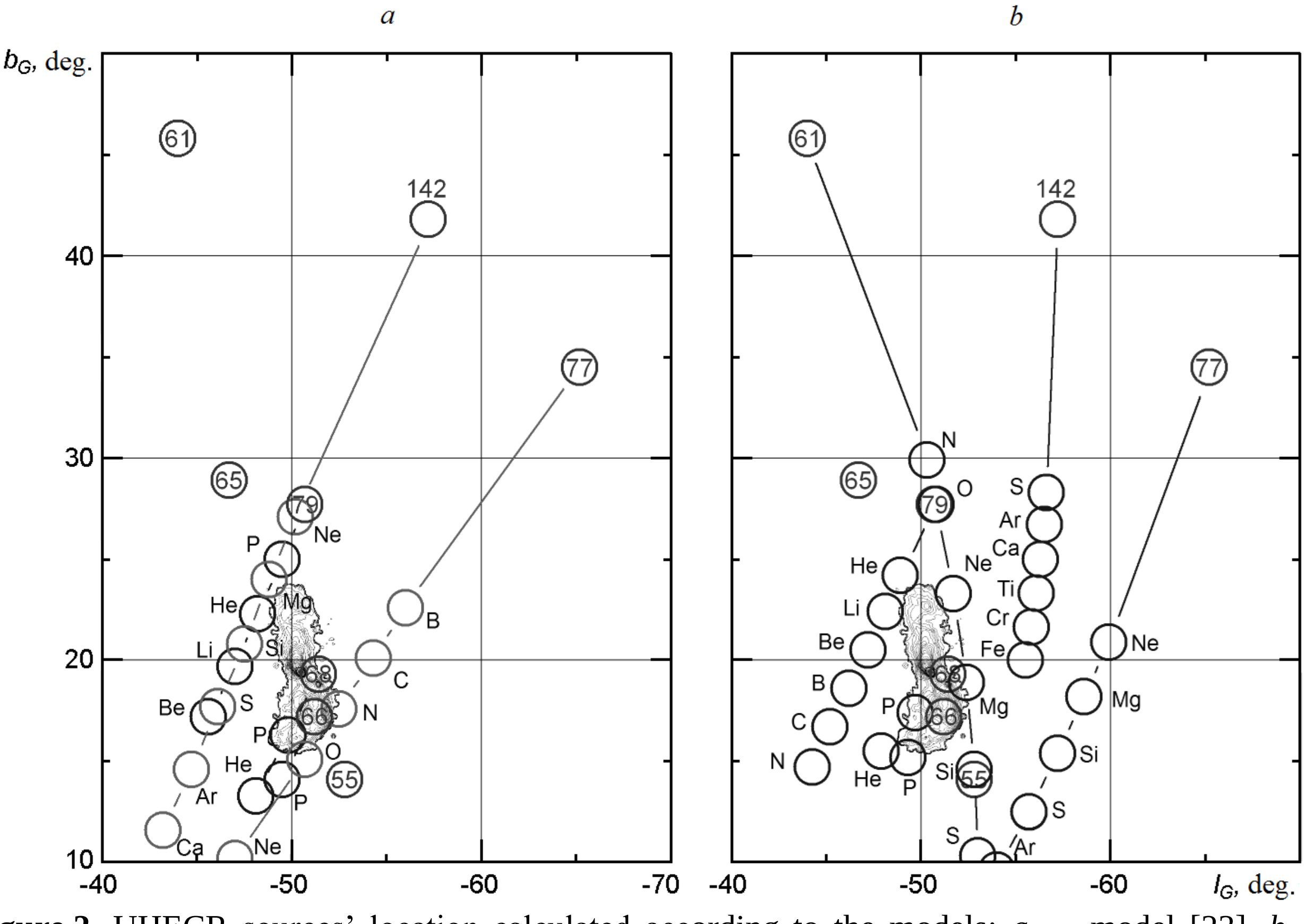

**Figure 2.** UHECR sources’ location calculated according to the models: *a* — model [23], *b* — model [21]

Overlapping of the circles corresponding to the calculated sources' location and the image of Centaurus A was chosen as the criterion for defining the correlation of the analyzed events and the said galaxy. Considering the Galactic field's random component as well as the extragalactic field widened the area of the source's possible localization without actually changing its location.

We have found out that out of all CR coming from Centaurus A's area only six can in fact originate in this galaxy – those with the energy of 61, 66, 68, 77, 79 and 142 EeV. Table 3 demonstrates chemical composition of the particles with the indicated energy. These particles correlate with Centaurus A following the two chosen models of the Galactic field and considering the impact of the magnetic field's various components (RegF – regular Galactic field; RmF – random Galactic field; EF – extragalactic field).

**Table 3.** UHECR's – Centaurus A correlation

| CR's energy, EeV | CR's chemical composition | | | | | |
|---|---|---|---|---|---|---|
| | model [23] | | | model [20] | | |
| | RegF | RegF + RmF | RegF + RmF + EF | RegF | RegF + RmF | RegF + RmF + EF |
| 142 | Mg | Mg — Ar | Ne — Ca | — | Ca — Fe | S — Fe |
| 79 | He | He — Li | p — Be | He — Li | He — C | He — N |
| 77 | N — O | C — Ne | B — Ne | — | Mg — Ar | Ne — Ar |
| 68 | p | p | p — He | p | p — He | p — He |
| 66 | — | p | p | p | p | p |
| 61 | — | — | — | Ne — Mg | O — S | N — S |
| CR's energy, EeV | CR's chemical composition | | | | | |
| | model [24] —ASS symmetry | | | model [24] —BSS symmetry | | |
| | RegF | RegF + RmF | RegF + RmF + EF | RegF | RegF + RmF | RegF + RmF + EF |
| 68 | p | p | p | — | — | p |
| 66 | p | p | p | — | — | p |

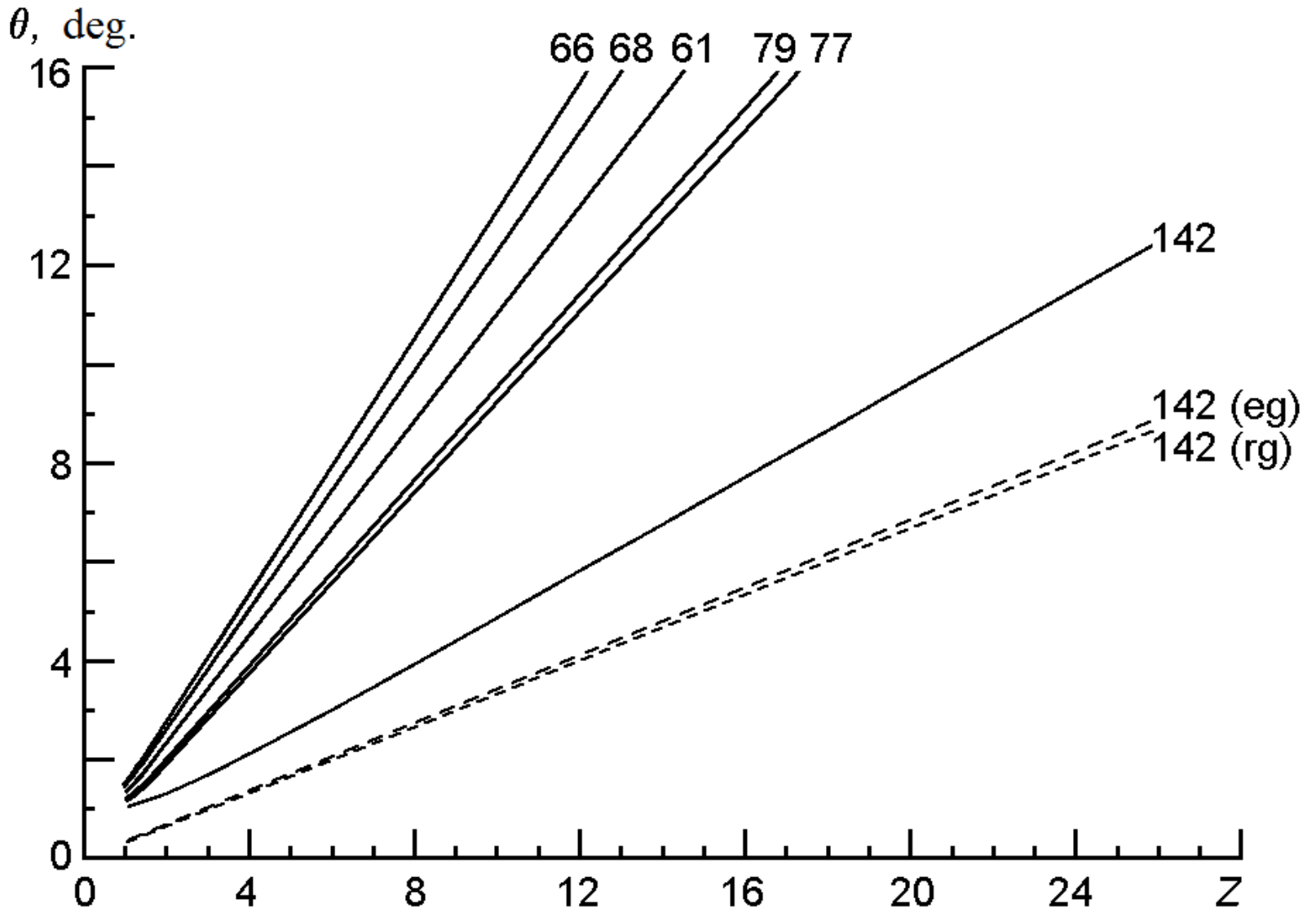

**Figure 3.** The size of CR's localization areas as influenced by random magnetic fields. Solid lines – the resulting deflection considers extragalactic and Galactic random fields as well as Auger equipment's error. Stroke lines – only deflection in extragalactic (eg) or galactic (rg) random fields are considered, CR having the energy of 142 EeV

Figure 3 demonstrates the size of CR's localization areas that correlate with Centaurus A, the latter being estimated in regard to possible deflections in both random Galactic and extragalactic magnetic fields.

## CONCLUSIONS

Centaurus A radio galaxy is treated off as the closest object that could be the source of some registered UHECR within the framework of present-day models of Galactic and extragalactic magnetic fields. Centaurus A may be the source of the events in its nearby region registered by the Auger observatory. Models [23] and [20] provide similar results. According to the calculations carried out on the basis of model [23] five events correlate with Centaurus A. When model [20] is employed we speak of six such events. The common tendency of shifting CR's chemical composition towards heavier nuclei at the boost of energy of the corresponding event is noted in all cases of possible correlation with the object under investigation. It is relevant for both models. The achieved result is in accord with the conclusions on UHECR's chemical composition made at the Auger observatory.

On the other hand, if the models suggested in [24] are concerned only two out of the studied CR (provided they are protons) correlate with Centaurus A. This result is supported by the conclusions made at the HiRes observatory.

Considering the impact of both Galactic and extragalactic fields may result into discovering a bigger number of nuclei correlating with Centaurus A. However in this case reliability of the events' connection to the Centaurus A source is diminished.

The source's comparative proximity to our Galaxy allows implementing the "back tracking" method for estimating Centaurus A's potential influence upon the observed UHECR flow, especially upon the events' arrival trajectories close to the direction towards this radio galaxy. Present-day models of the Galactic magnetic field agree with the fact that a number of events registered by the Auger observatory correspond to the UHECR accelerated in Centaurus A. Observing larger number of events from Centaurus A region as well as improving the models of Galactic and extragalactic magnetic fields may be crucial for proving the possibility of UHECR's acceleration in galaxies of Centaurus A type.